\documentclass[reprint,onecolumn,amsmath,amssymb,aps,
]{revtex4}

\usepackage{epsfig,bm,epsf,graphics}
\usepackage{graphicx}% Include figure files
\usepackage{dcolumn}% Align table columns on decimal point
\usepackage{bm}% bold math
\usepackage{lscape}
\usepackage{xcolor}

\begin{document}
%\draft
\preprint{APS/123-QED}

\title{IS THE HYPERSCALING RELATION VIOLATED BELOW THE UPPER CRITICAL DIMENSION IN SOME PARTICULAR CASES
?}% Force line breaks with \\

\author{Hung T. Diep$^{(a)}$ \footnote{diep@cyu.fr, corresponding author} and Van-Thanh Ngo$^{(b)}$\footnote{nvthanh@vast.vn}}
\address{$^{(a)}$ Laboratoire de Physique Th\'eorique et Mod\'elisation,
CY Cergy Paris Universit\'e, CNRS,
UMR 8089\\
2 Avenue Adolphe Chauvin, 95302 Cergy-Pontoise, France\\
$^{(b)}$ Institute of Scientific Data and Information, Vietnam Academy of Science and Technology (VAST)\\ 18 Hoang Quoc Viet, Hanoi 10000, Vietnam}

\date{\today}% It is always \today, today,


\begin{abstract}

In this review, we show our results with new interpretation on the critical exponents of  thin films obtained by  high-performance multi-histogram Monte Carlo simulations. The film thickness $N_z$ consists of a  few layers up to a dozen of  layers in the $z$ direction.  The free boundary condition is applied in this  direction while in the $xy$ plane periodic boundary conditions are used.  Large $xy$ plane sizes are  used for finite-size scaling.  The Ising model is studied with nearest-neighbor (NN) interaction. When $N_z=1$, namely the two-dimensional (2D) system, we find  the  critical exponents given by the renormalization group.  While, for $N_z>1$, the critical exponents calculated with the high-precision multi-histogram technique show that they deviate slightly but systematically from the 2D values.   If we argue that as long as the thickness $N_z$ is small enough, the correlation length in the $z$ direction is finite, it does not affect the nature of the phase transition, namely it remains in the 2D universality class. This argument is in contradiction with our numerical results which show a systematic deviation from 2D values.  If we use these values of critical exponents in the hyperscaling relation with $d=2$, then the hyperscaling relation is violated.  However, if we use the hyperscaling relation and the critical exponents obtained for $N_z>1$ to calculate  the dimension of the system, we find the system dimension slightly larger than 2.  This can be viewed as an "effective" dimension.  More discussion is given  in the paper.  We also show the cross-over between the first- and second-order transition while varying the film thickness in an antiferromagnetic FCC Ising frustrated thin film.  In addition, we will show evidence that when a 2D system has two order parameters of different symmetries with a single transition, the critical exponents are new, suggesting a universality class of coupled two-symmetry breakings.  In this case, the 2D hyperscaling does not hold. Another case is the  3D Ising model  coupled to the lattice vibration: the critical exponents deviate from the 3D Ising ones, the results suggest the violation of the hyperscaling. 
\vspace{0.5cm}
\begin{description}
\item[PACS numbers: ]
\item[Keywords: Critical Exponents; Thin Films; Hyperscaling Relations; Effective Dimension; ]
\item[\hspace{2cm}Multi-Histogram Monte Carlo Technique.]
\end{description}
\end{abstract}

\maketitle

\section{Introduction}
The study of phase transitions is one of the most important tasks in theory, in experiments and in computer simulations since the second half of the 20th century.  The main reason is that, if we know the characteristics
of a phase transition, we can understand the interaction mechanisms  lying behind the transition and we can deduce various physical quantities.  Therefore, comparisons between theories, experiments and computer simulattions are necessary  in order to obtain conclusions.  While comparison with experiments is always a challenge because real materials may contain  ill-controlled elements such as dislocations, defects and impurities, comparisons between theories and  simulations are in most cases possible.   This is the purpose of the present short review based on our own works in the past years (see our works cited in \cite{HTD2025,HTD2021}). 

Let us recall that the phase transition was first studied  by the mean-field approximation with several improved versions in the 40s (see the textbook \cite{DiepTM} where these methods are shown and commented). These were followed by exact-solution methods in two dimensions (2D) such as the Ising model, Potts models and vertex models (see the book by R. Baxter \cite{Baxter}). The break-through in general dimensions come in 1970 with the formulation of the renormalization group by K. G. Wilson \cite{Wilson1,Wilson2} followed by  investigations on the finite-size scaling analysis and their validity \cite{Fisher1,Fisher2,Cardy}. 
As known, there are  six critical exponents and there are four relations between them. One is called 'hyperscaling relation' connected  with the space dimension $d$.  There have been investigations on the progress made on the scaling and
hyperscaling relations above the  upper critical dimension $d_u=4$.  The reader is referred, for example, to Ref.\cite{Kenna} and the review by  Young \cite{Young} for a recall of the demonstration of these relations (see also  the review by Honchar et al. \cite{Honchar24}).  
It should be added that this violation is explained in the framework of Renormalization-Group
theory by the existence of a dangerous irrelevant variable \cite{Luijten}. 
The main question  is whether or not the hyperscaling is violated for dimension $d$  larger than the upper critical dimension $d_u=4$.  In the present paper, however, we show that the question of the violation of the hyperscaling is also posed in $d < 4$ in some specific cases  that we will present in this paper. It should be mentioned that Renormalization-Group calculations conclude that the hyperscaling relation should hold for thin films in the case of large-$N$ limit \cite{OConnor}. But in the present paper, we consider thin films of Ising spins ($N=1)$ in section \ref{itf}, so the result of this paper does not apply.

The first case studied in the present paper concerns the  critical exponents obtained by using the highly-precise multi-histogram Monte Carlo (MC) technique \cite{Ferrenberg1,Ferrenberg2,Bunker} for a thin film of simple cubic structure with Ising spin model. The film surface $L^2$ ($xy$ plane) is very large, up to $L^2=160^2$ lattice sites for some cases, with periodic boundary conditions, while the film thickness $N_z$ goes from one layer (2D)  to 13 layers with free boundary conditions. Finite-size scaling (FSS) has been used with varying $L$ to calculate the critical exponents.  These results have been published in Ref. \cite{Ngo2009} but the aspect of the violation of the hyperscaling relation has not been discussed. In the light of the violation of the hyperscaling relation for $d\geq d_u$ \cite{Kenna,Young}, we revise the interpretation of our results. In addition, we review some other cases that we have investigated.

Except the case $N_z=1$, the question of the dimension of the system naturally arises.  For Capehart and Fisher \cite{Fisher}, there is a cross-over from 2D to 3D when $N_z$ is increased. So, how should one use the hyperscaling relation, namely with which dimension ? We will discuss this in this paper.

This  review is organized as follows. 
Section \ref{itf} is devoted to the case of thin films mentioned above where the multi-histogram technique is recalled. Section \ref{Results} shows the results of critical exponents obtained with the FSS.
Section \ref{Crossover} reviews the case of a cross-over between the first- and second-order transitions when the film thickness of a frustrated antiferromagnetic FCC film is decreased. 
Section \ref{Other} is devoted to some other cases where the hyperscaling relation is violated or seems to be violated.
Concluding remarks are given in section \ref{Concl}.

\section{Critical Behavior of Thin Films}\label{itf}
\subsection{Model} 
We consider a thin film composing of $N_z$ layers of $xy$ square lattices stacking in the $z$ direction. The $xy$ plane has the dimension $L\times L$ where $L$ is the linear dimension.  We use
a simple Hamiltonian which consists of Ising spins occupying the lattice sites. The spins interact with each other via an exchange interaction $J$ between nearest neighbors (NN). We assume $J$ to be uniform everywhere including between surface spins. The Hamiltonian is written as
\begin{equation}\label{eqn:hamil1}
{\cal H}=-J\sum_{i,j} \sigma_i \ \sigma_j .
\end{equation}
In the simulations which will be shown below, we take $N_z$ from 1 to 13 and $L=20-80$ (for some cases $L$ is up to 160). Before showing our results, let us summarize the multi-histogram technique in the following.

\subsection{Multi-histogram Technique}\label{multihistogram}
The principle of a single histogram MC method is to  record the energy histogram $H(E)$ collected at a temperature $T$ as  close as possible to the transition temperature obtained from the standard MC simulation. From $H(E)$ one can calculate thermodynamic quantities using formulas of the canonical distribution. The error depends on how far the chosen temperature is from the transition  temperature $T_c$.   Multi-histogram technique helps overcome this uncertainty.  We know that with a finite-size system the maximum of the specific heat $C_v$ and the maximum of the susceptibility $\chi$ do not occur at the same temperature. In the multi-histogram technique, we take another temperature close to the maximum of $C_v$ and  the maximum of $\chi$ to calculate by histogram technique the new temperatures of the maxima of $C_v$ and $\chi$. We repeat again this procedure  for  8 temperatures: such an iteration procedure improves the positions of the maxima of $C_v$ and $\chi$ at the system size $(L,N_z)$.   
The reader is referred to Refs. \cite{Ferrenberg2,Bunker} for more technical details.   
Note that, as in the single histogram technique,  thermal physical quantities are calculated as
continuous functions of $T$ which allow for the precise determination of the peak positions and the peak heights of $C_v$ and $\chi$ for a given system size. This permits making the finite-size scaling with precision. In addition, since we take many temperatures in the transition region, the results obtained by multi-histogram method  are valid for a wider range of temperature, unlike a single histogram technique with the results  valid only in a very small temperature region, typically $[T-T_c(\infty)]/T_c(\infty)\simeq \pm 1\%$.

The multiple histogram technique is known to reproduce with very
high accuracy the critical exponents of second order phase
transitions.\cite{Ferrenberg1,Ferrenberg2,Bunker}
The overall probability distribution\cite{Ferrenberg2} at
temperature $T$ obtained from $n$ independent simulations, each
with $N_j$ configurations, is given by
\begin{equation}
P(E,T)=\frac{\sum_{i=1}^n H_i(E)\exp[E/k_BT]}{\sum_{j=1}^n
N_j\exp[E/k_BT_j-f_j]}, \label{eq:mhp}
\end{equation}
where
\begin{equation}
\exp[f_i]=\sum_{E}P(E,T_i). \label{eq:mhfn}
\end{equation}
The thermal average of a physical quantity $A$ is then calculated
by
\begin{equation}
\langle A(T)\rangle=\sum_E A\,P(E,T)/z(T),
\end{equation}
in which
\begin{equation}
z(T)=\sum_E P(E,T).
\end{equation}
Thermal averages of physical quantities are thus calculated as
continuous functions of $T$, now the results should be valid over
a much wider range of temperature than for any single histogram.
In practice, we use first the standard MC simulations to localize
for each size the transition temperatures $T^E_0(L)$ for specific
heat and $T^m_0(L)$ for susceptibility. The equilibrating time is
from 200000 to 400000 MC steps/spin and the averaging time is from
500000 to 1000000 MC steps/spin. Next, we make histograms at $8$
different temperatures $T_j(L)$ around the transition temperatures
$T^{E,m}_0(L)$ with 2 millions MC steps/spin, after discarding 1
millions MC steps/spin for equilibrating. Finally, we make again
histograms at $8$ different temperatures around the new transition
temperatures $T^{E,m}_0(L)$ with $2\times 10^6$ and $4\times 10^6$
MC steps/spin for equilibrating and averaging time, respectively.
Such an iteration procedure gives extremely good results for
systems studied so far.  Errors shown in the following have been
estimated using statistical errors, which are very small thanks to
our multiple histogram procedure, and fitting errors given by
fitting software.

In MC simulations,  we calculate the thermal averages of 

\noindent - the magnetization$\langle M\rangle$
\begin{equation}
\langle M\rangle=\frac{1}{L^2N_z}\langle\sum_i\sigma_i\rangle,
\end{equation}
- the total energy $\langle E\rangle$, 
\begin{equation}
\langle E\rangle=\langle\cal{H}\rangle,
\end{equation}
- the heat capacity $C_v$, 
\begin{equation}
C_v=\frac{1}{k_BT^2}\left(\langle E^2\rangle-\langle E\rangle^2\right),
\end{equation}
- the susceptibility $\chi$, 
\begin{equation}
\chi=\frac{1}{k_BT}\left(\langle M^2\rangle-\langle M\rangle^2\right),
\end{equation}
- the Binder  energy cumulant $U$,
\begin{equation}
U=1-\frac{\langle E^4\rangle}{3\langle E^2\rangle^2},
\end{equation}
-$n^{th}$ order cumulant of the order parameter $V_n$ for $n=1$
and 2: 
\begin{equation}
V_n=\frac{\partial\ln{M^n}}{\partial(1/k_BT)} =\langle
E\rangle-\frac{\langle M^nE\rangle}{\langle M^n\rangle}.
\end{equation}

Note that if we scale with $L$, we have the following relations, since $N_z$ is fixed \cite{Young,Ferrenberg2,Binder1985}: 
\begin{equation}\label{V1}
V_1^{\max}\propto L^{1/\nu}, \hspace{1cm} V_2^{\max}\propto
L^{1/\nu},
\end{equation}
\begin{equation}
C_v^{\max}=C_0+C_1L^{\alpha/\nu}\label{Cv}
\end{equation}
and
\begin{equation}\label{chis}
\chi^{\max}\propto L^{\gamma/\nu},
\end{equation}
at their respective 'transition' temperatures $T_c(L)$, and

\begin{equation}\label{CU}
U=U[T_c(\infty)]+AL^{-\alpha/\nu},
\end{equation}
\begin{equation}
M_{T_c(\infty)}\propto L^{-\beta/\nu}\label{MB},
\end{equation}
and
\begin{equation}\label{TCL}
T_c(L)=T_c(L=\infty)+C_AL^{-1/\nu}.
\end{equation}
In the above relations,  $A$, $C_0$, $C_1$ and $C_A$ are constants.  The exponent $\nu$ can be calculated
by Eqs. (\ref{V1}). For $L$  large enough, $V_1^{\max}$ and $V_2^{\max}$ should give the same $\nu$ as seen below.  Then, from Eq. (\ref{TCL}) we estimate $T_c(L=\infty)$. Using this, we calculate $\alpha$ and $\beta$ from Eqs. (\ref{CU}) and (\ref{MB}).  We will check the  Rushbrooke inequality $\alpha+2\beta+\gamma\geq 2$ and the hyperscaling relation $d\nu=2-\alpha$ in the following.

To ensure that Eqs. (\ref{V1})-(\ref{TCL}) are used for $L$ large enough,  one uses the following corrections to scaling of the form
\begin{eqnarray}
\chi^{\max}& =& B_1L^{\gamma/\nu}(1+B_2L^{-\omega})\label{chic},\\
V_n^{\max}&=& D_1L^{1/\nu}(1+D_2L^{-\omega})\label{V1c},
\end{eqnarray}
$B_1$, $B_2$, $D_1$  and $D_2$ being constants and $\omega$  a
correction exponent.\cite{Ferrenberg3} Normally, if $L$ is large enough, the corrections are very small.  With today's computer memory capacity, large $L$ is often used as seen below. The scaling corrections are thus extremely small, they do not therefore alter the results using Eqs. (\ref{V1})-(\ref{TCL}).

\section{Results}\label{Results}
Most of the results shown below have been published by us in \cite{Ngo2009}. However, in light of the new interpretation on the violation of the hyperscaling relation, we review them here to show that there is a possibility that the hyperscaling relation $d\nu=2-\alpha$ is violated below $d_u$: as mentioned in the Introduction, in the case of thin films where the thickness is finite and small, the hyperscaling relation is satisfied only when $d$ has a value between 2 and 3.   This non-integer dimension is called by Capehart and Fisher \cite{Fisher} "cross-over dimension".  To our knowledge, $d$ in the hyperscaling relation is the space dimension, it is 2, 3, ... (integer), it cannot be between 2 and 3. We are convinced that the hyperscaling relation may not be valid in the case of  thin films shown below and in other particular cases shown in  section \ref{Other}.

\begin{figure} 
\centering
\includegraphics[width=3.2 in]{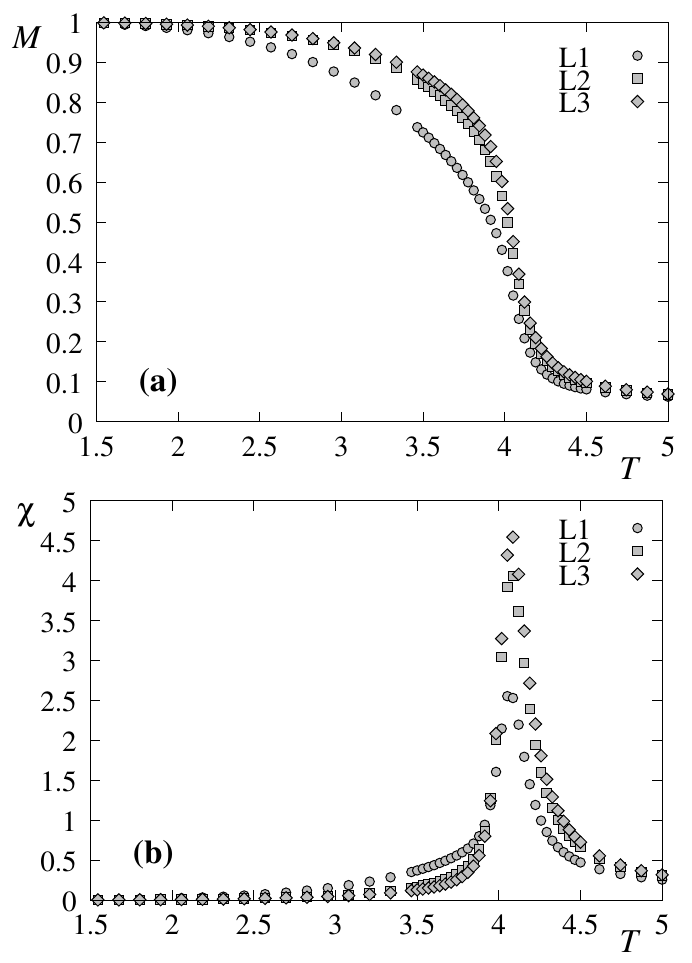}
\caption{ (a) Layer
magnetizations of layer 1 (denoted by $L_1$), layer 2 (denoted by $L_2$) and layer 3 (denoted  by $L_3$), (b) Layer susceptibilities, as functions of $T$ with $N_z=5$ and $L=24$.} \label{fig:N24Z5M}
\end{figure}

\begin{figure}
\centering
\includegraphics[width=3.2 in]{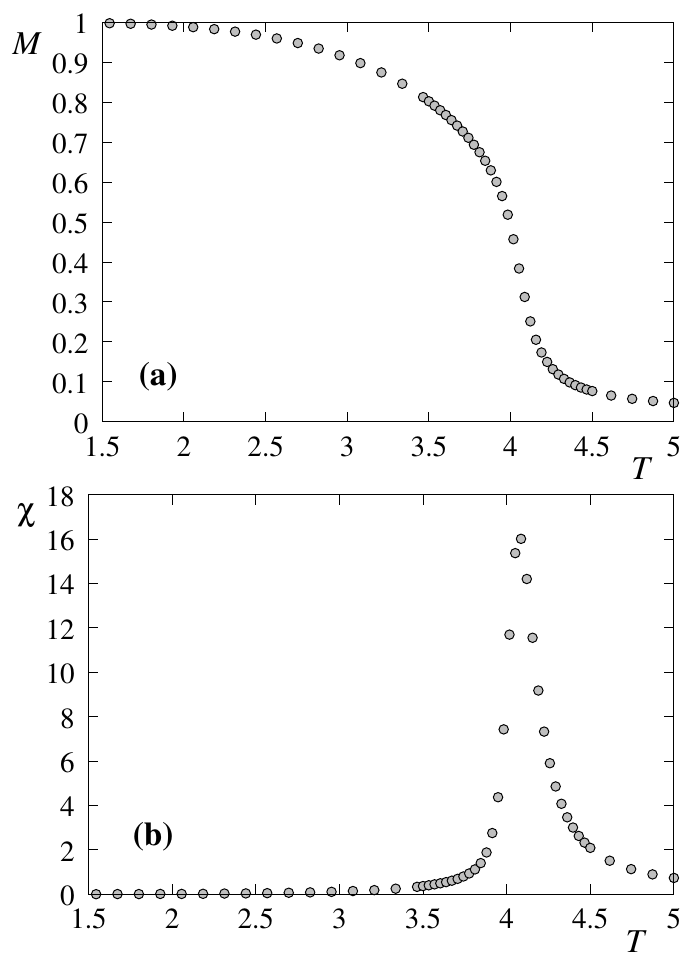}
\caption{(a) Total
magnetization, (b) Total susceptibility,  versus $T$ with
$N_z=5$ and $L=24$.} \label{fig:N24Z5MT}
\end{figure}

First, we show the layer magnetizations and
their corresponding susceptibilities of the first three layers in the case where $N_z=5, L=24$ in Fig. \ref{fig:N24Z5M}.  The layer susceptibilities have their peaks  at the same temperature,
indicating a single transition.  We note that the magnetization is lowest at the surface and increases while going to the interior. This is known a long time ago
by the Green's function method \cite{DiepTM,diep79,diep81} and by more recent works on thin films with competing interactions \cite{Diep2015} and films with Dzyaloshinskii-Moriya interaction \cite{Sahbi2017,Sharafullin2019}.  Physically, the surface spins have smaller local field
due to the lack of neighbors, so thermal fluctuations will reduce
more easily the surface magnetization with respect to the interior
ones. 
%Fig1
We plot in Fig. \ref{fig:N24Z5MT} the total magnetization and
the total susceptibility. The latter shows only one peak, signature of a single transition.  This justifies the study shown below on the criticality of the film transition.
%Fig2

\subsection{Finite size scaling}
Let us show our finite-size scaling (FSS) with $L$ varying from 20 to 80. For $N_z=3$ we use $L$ up to 160 to evaluate the corrections to scaling. 
The technical details are described in the subsection  \ref{multihistogram} above.
Note that the multi-histograms at 8 temperatures in  the critical region have been performed iteratively many times.  Such an iteration procedure gives extremely good results.  Errors shown in the following have been
estimated using statistical errors, which are very small thanks to
our multiple histogram procedure, and fitting errors given by
fitting software. Note that the errors are at the 3rd digit as seen in Table \ref{tab:criexp}, they are smaller than the size of the data points shown in the figures.

\subsection{Critical exponents obtained with finite-size scaling}

\begin{figure}
\centering
\includegraphics[width=4 in]{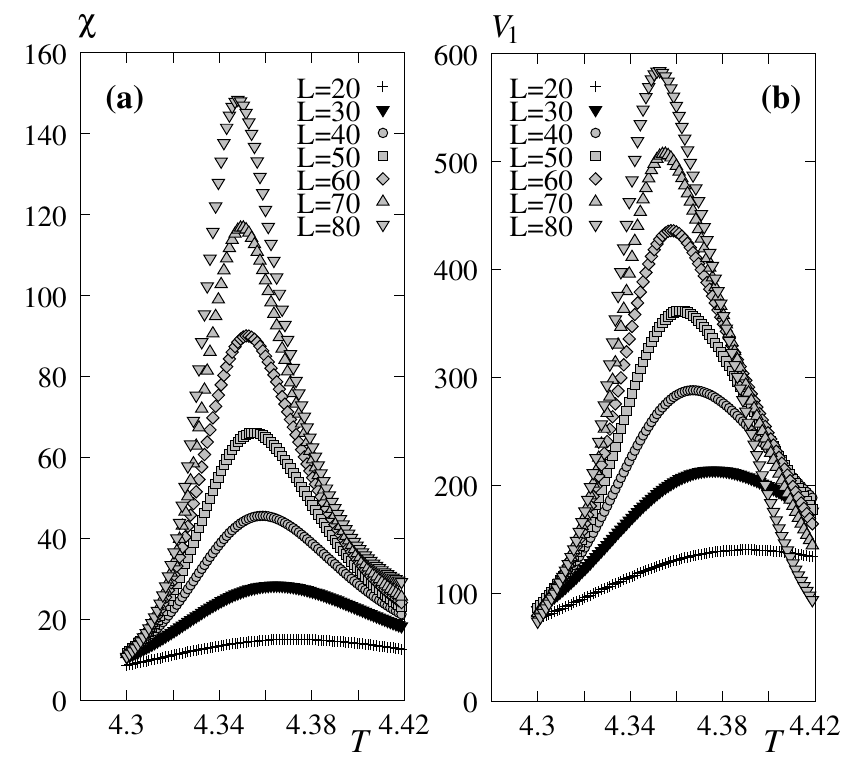}
\caption{(a) Susceptibility and (b) $V_1$, as functions of $T$ for
 $L=20,30,...,80$ with $N_z=11$, obtained by multiple histogram
technique. Note the strong dependence of the peak height on $L$.} \label{fig:ISSVZ11}
\end{figure}

We show first the peak height of the susceptibility and the maximum of $V_1$ as functions of $T$ for varying $L$ from 20 to 80 in Fig. \ref{fig:ISSVZ11}.  

%Fig3

\begin{figure}
\centering
\includegraphics[width=4 in]{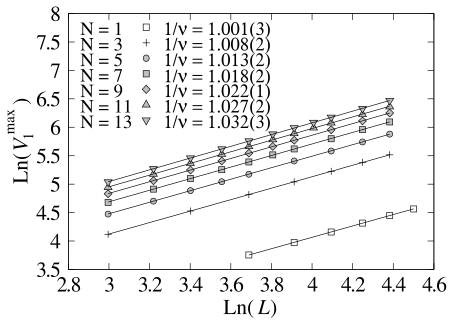}
\caption{ Maximum
of $V_1$  versus $ L$ in the $\ln-\ln$
scale for various $N_z$.  The  slopes give $1/ \nu$ [see Eq. (12)].  These values are indicated on the figure. } \label{fig:NUL}
\end{figure}

\begin{figure}
\centering
\includegraphics[width=4 in]{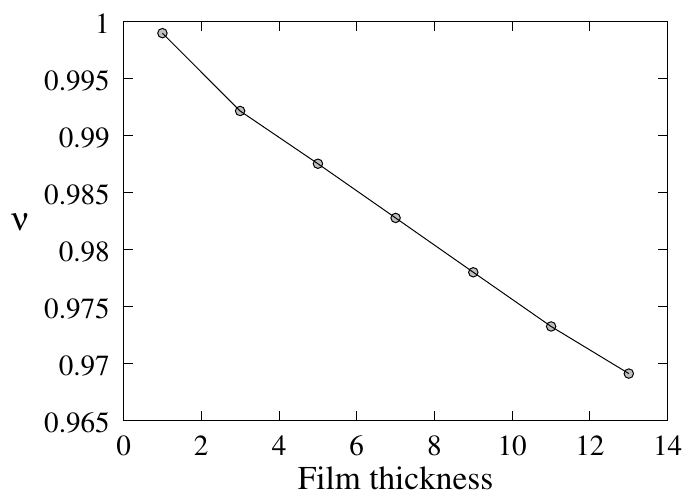}
\caption{Exponent $\nu$ versus $N_z$. Note that $\nu(2D)=1$ and $\nu(3D)=0.62997$.} \label{fig:NUZ}
\end{figure}

Note that the results shown below are obtained using $T_c(L=\infty, N_z)$  as described earlier below Eq. (\ref{TCL}). We show in Fig. \ref{fig:NUL} $V_1^{max}$ versus $L$ in the $ln-ln$ scale for $N_Z=1,3,...,13$.  The slope of a given straight line gives $1/\nu$ of the corresponding $N_z$. 
%Fig4
Exponent $\nu$ obtained from Fig. \ref{fig:NUL}  is  shown as a function of $N_z$ in Fig. \ref{fig:NUZ} 

%Fig5

\begin{figure}
\centering
\includegraphics[width=4 in]{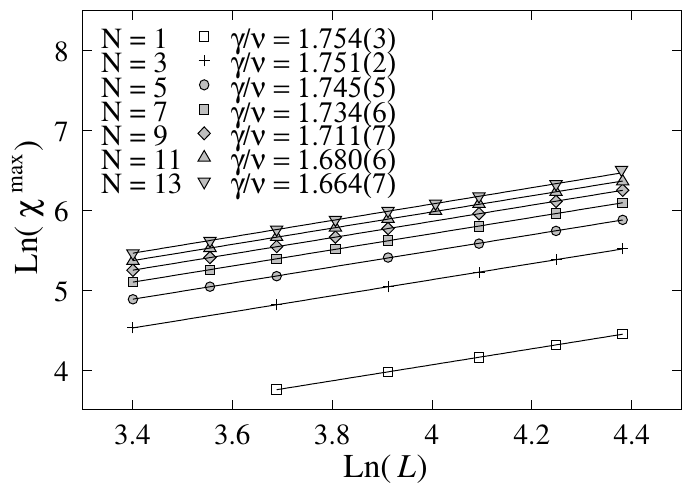}
\caption{Maximum
of susceptibility versus $L$ in the $\ln-\ln$ scale. The  slopes give $\gamma/\nu$ 
indicated on the figure.} \label{fig:ISGAML}. 
\end{figure}

\begin{figure}
\centering
\includegraphics[width=4 in]{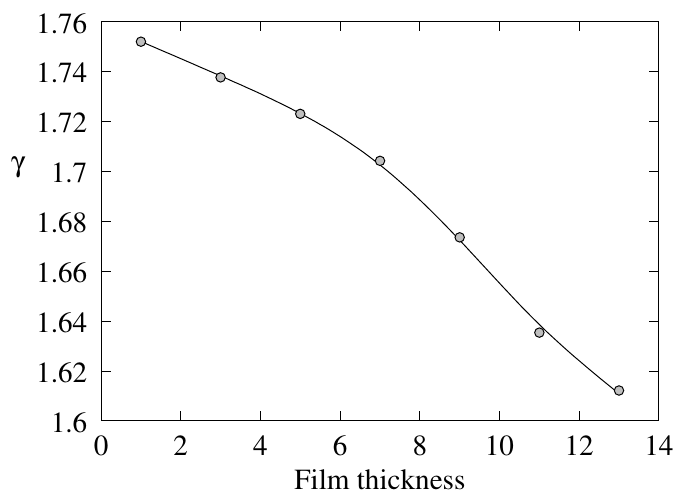}
\caption{Exponent $\gamma$ versus $N_z$. Note that $\gamma(2D)=1.75$ and $\gamma(3D)=1.23707$.} \label{fig:ISGAMZ}
\end{figure}

\begin{figure}
\centering
\includegraphics[width=4 in]{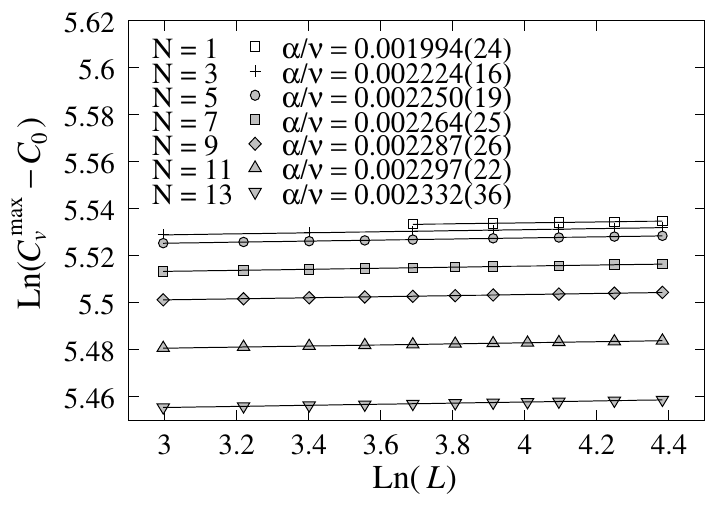}
\caption{$\ln
 (C_v^{\max}-C_0)$ versus $\ln L$ for
$N_z=1,3,5,...,13$.  The slope gives $\alpha/\nu$  (see Eq. 
\ref{Cv}) indicated on the figure.  Note that $\alpha/\nu(2D)=0$ and $\alpha/\nu(3D)=0.110087/0.62997=0.174749.$} \label{fig:ALPHAL}
\end{figure}

To show the precision
of our method, we give here the results of $N_z=1$. For $N_z =1$,
we have $1/\nu =1.0010 \pm 0.0028$ which yields $\nu = 0.9990\pm
0.0031$ and $\gamma/\nu = 1.7537 \pm 0.0034$ and  $\gamma =
1.7520\pm 0.0062$. These results are in excellent agreement with
the exact results $\nu_{2D}=1$ and $\gamma_{2D}=1.75$.  The very
high precision of our method is thus verified in the rather modest
range of the system sizes $L=20-80$ used in the present work. Note
that the result of Ref.\cite{Schilbe} gave $\nu=0.96\pm0.05$
for $N_z=1$ which is very far from the exact value.
The results of $\gamma/\nu$  are shown  in Figs. \ref{fig:ISGAML} and \ref{fig:ISGAMZ}. The results of $\alpha/\nu$  are shown  in Fig. \ref{fig:ALPHAL}. These critical exponents are, although close to those of 2D, show a systematic deviation from the 2D case with increasing $N_z$.

%Fig6

%Fig7

%Fig8

\subsection{Corrections to scaling}\label{correction}
\begin{figure}
\centering
\includegraphics[width=4 in]{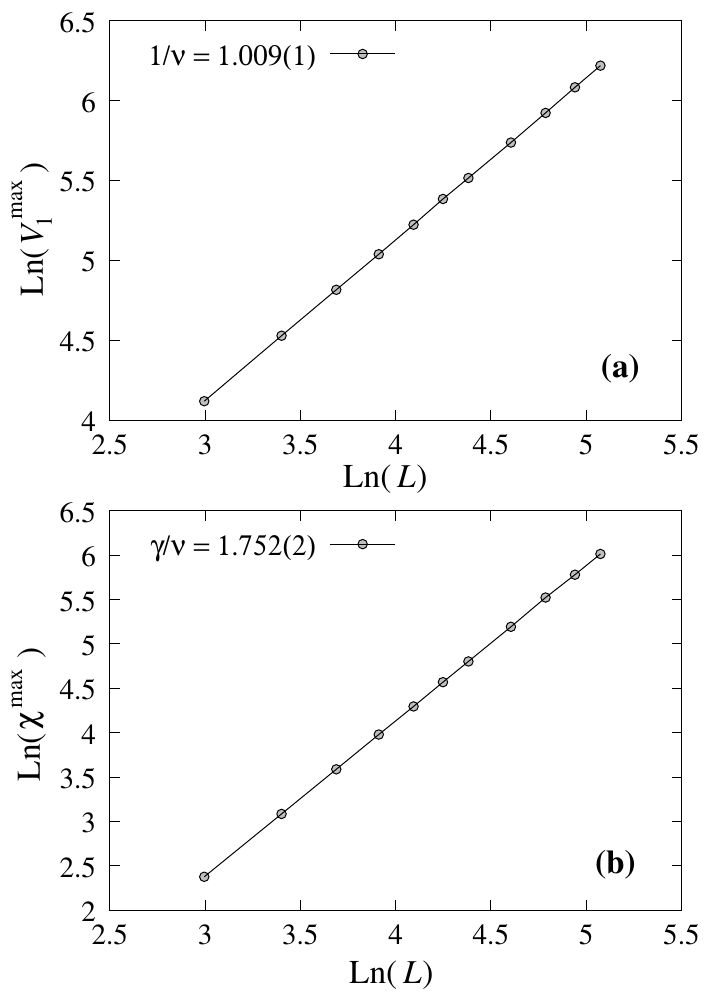}
\caption{(a)
$V_1^{max}$ and (b) $\chi^{max}$ vs $L$ up to 160 with $N_z=3$.}
\label{fig:GNL160}
\end{figure}

Let us touch upon the question of corrections to scaling mentioned earlier.
We show now that  the corrections to scaling are very small.
We consider here the effects of larger $L$ and of the correction
to scaling   for $N_z=1,3,...,13$. 
The results indicate that larger $L$ does not change the results
shown above. Figure \ref{fig:GNL160}(a) displays the maximum of
$V_1$ as a function of $L$ up to 160. Using Eq. (\ref{V1}), i.e.
without correction to scaling, we obtain $1/\nu=1.009\pm0.001$
which is to be compared to $1/\nu=1.008\pm0.002$ using $L$ up to
80. The change is therefore insignificant because it is at the
third decimal i. e. at the error level.  The same is observed for
$\gamma/\nu$ as shown in Fig. \ref{fig:GNL160}(b): $ \gamma/\nu =
1.752\pm 0.002$ using $L$ up to 160 instead of $ \gamma/\nu =
1.751\pm 0.002$ using $L$ up to 80.

%Fig9

%Fig10
\begin{figure}
\centering
\includegraphics[width=4 in]{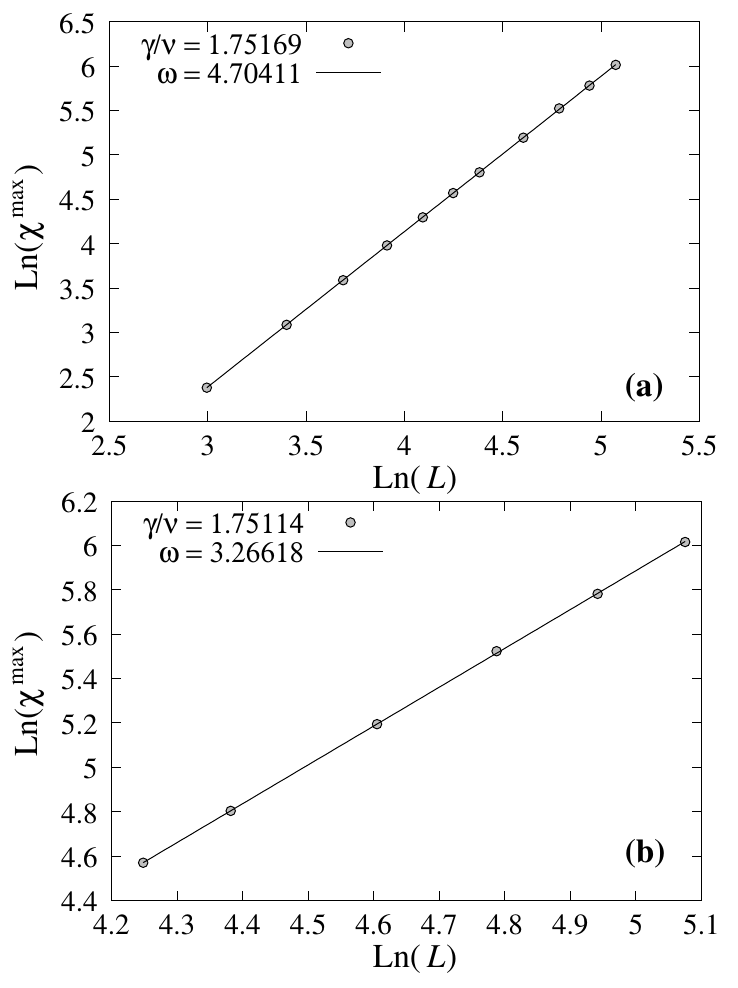}
\caption{$\chi^{max}$ vs $L$ (a) from $20$ up to 160  (b) from
$70$ up to 160, for $N_z=3$.} \label{fig:CORRECT}
\end{figure}
Now, let us allow for correction to scaling, i. e. we use
Eq.(\ref{chic}) instead of Eq. (\ref{chis}) for fitting. We obtain
the following values:  $\gamma/\nu=1.751\pm0.002$, $B_1= 0.05676$,
$B_2=1.57554 $, $\omega=3.26618$ if we use $L$ = 70 to 160 (see Fig.
\ref{fig:CORRECT}). The value of $\gamma/\nu$ in the case of no
scaling correction is $1.752\pm0.002$. Therefore, we can conclude
that this correction is insignificant. The large value of $\omega$
explains the smallness of the correction.

\begin{figure}
\centering
\includegraphics[width=4 in]{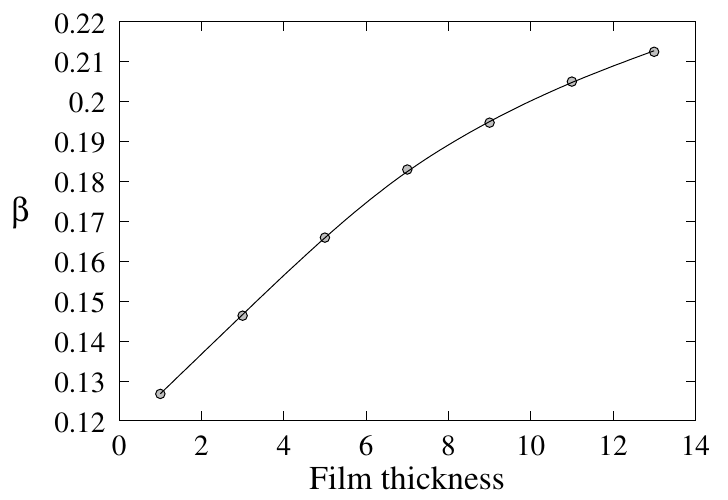}
\caption{Exponent $\beta$ as a function of the film thickness. } \label{BETA}
\end{figure}

For $\beta$, using Eq. (\ref{MB})  we calculate $\beta/\nu$ for each thickness $N_z$. The results are precise. For example for $N_z$=1, we obtained $\beta/\nu = 0.1268 \pm 0.0022$ which yields
$\beta = 0.1266\pm 0.0049$ which  is in agreement within errors with the exact result $\beta=1.25$. We show in Fig. \ref{BETA} the exponent $\beta$ versus $N_z$.

%Fig11

\subsection{Summary of our results}
We summarize our results in Table \ref{tab:criexp}.  Note that if we use $\nu$ and $\alpha$ for a given $N_z$,  
except the case $N_z=1 (d=2)$, the hyperscaling relation $d\nu=2-\alpha$ is violated if $d=2$. This relation is obeyed if the dimension is replaced by an "effective dimension" $d_{\mbox{eff}}$ listed in Table \ref{tab:criexp} which is a little bit larger than 2. This means that the 2D critical behavior dominates in thin films as expected ($d_{\mbox{eff}}$ varies from 2 to 2.0613 for $N_z=1$ to 13). The last column shows the critical temperature at $L=\infty$ obtained by using Eq. (\ref{TCL}).
%\begin{landscape}
\begin{table*}
% \centering
   \begin{tabular}{| r | c | c | c | c | c | c |}
    \hline
    $N_z$ & $\nu$ & $\gamma$ & $\alpha$ & $\beta$ & $d_{\mathrm{eff}}$ &
$T_c(L=\infty,N_z)$ \\
\hline 1 & $0.9990 \pm 0.0028$ & $1.7520 \pm 0.0062$ & $0.00199
\pm 0.00279$ & $0.1266 \pm 0.0049$ & $2.0000 \pm 0.0028$ &
$2.2699\pm
0.0005$ \\
3 & $0.9922 \pm 0.0019$ & $1.7377 \pm 0.0035$ & $0.00222 \pm
0.00192$ & $0.1452 \pm 0.0040$ & $2.0135 \pm 0.0019$ & $3.6365 \pm
0.0024$ \\
5 & $0.9876 \pm 0.0023$ & $1.7230 \pm 0.0069$ & $0.00222 \pm
0.00234$ & $0.1639 \pm 0.0051$ & $2.0230 \pm 0.0023$ & $4.0234 \pm
0.0028$ \\
7 & $0.9828 \pm 0.0024$ & $1.7042 \pm 0.0087$ & $0.00223 \pm
0.00238$ & $0.1798 \pm 0.0069$ & $2.0328 \pm 0.0024$ & $4.1939 \pm
0.0032$ \\
9 & $0.9780 \pm 0.0016$ & $1.6736 \pm 0.0084$ & $0.00224 \pm
0.00161$ & $0.1904 \pm 0.0071$ & $2.0426 \pm 0.0016$ & $4.2859 \pm
0.0022$ \\
11& $0.9733 \pm 0.0025$ & $1.6354 \pm 0.0083$ & $0.00224 \pm
0.00256$ & $0.1995 \pm 0.0088$ & $2.0526 \pm 0.0026$ & $4.3418 \pm
0.0032$ \\
13& $0.9692 \pm 0.0026$ & $1.6122 \pm 0.0102$ & $0.00226 \pm
0.00268$ & $0.2059 \pm 0.0092$ & $2.0613 \pm 0.0027$ & $4.3792 \pm
0.0034$ \\
    \hline
\end{tabular}
\vspace{0.3cm}
 \caption{Critical exponents obtained by multti-histogram technique. The effective dimension and critical temperature $T_c(L=\infty,N_z)$ are listed in the last two columns. See text for the definition of  the effective dimension $d_{\mbox{eff}}$.}\label{tab:criexp}
\end{table*}
%\end{landscape}
%\end{sidewaystable}
%\end{widetext}
%%%%%%%%%%%%%%%%%%

\subsection{Discussion}
\begin{figure}
\centering
\includegraphics[width=4 in]{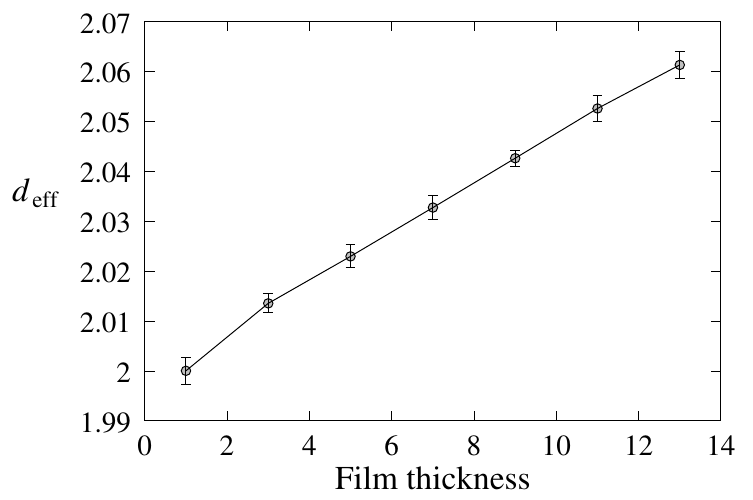}
\caption{Effective dimension $d_{\mbox{eff}}$ of thin film defined by $d_{\mbox{eff}}\nu=2-\alpha$, as a function of thickness. See text for comments.} \label{de}
\end{figure}

Let us show in  Fig. \ref{de} the effective dimension versus $N_z$.   As seen, though $d_{\mbox{eff}}$  deviates systematically from $d=2$, its values are however very close to 2. This means that the 2D character is dominant even at $N_z=13$. 

%Fig12 
\begin{figure}
\centering
\includegraphics[width=4 in]{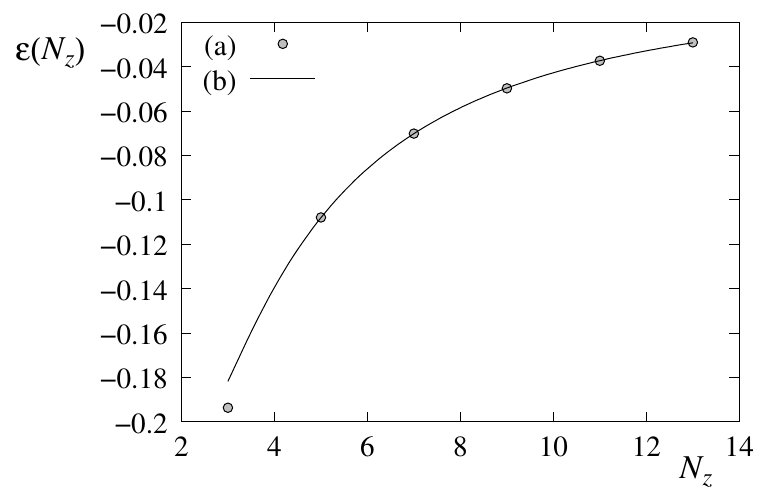}
\caption{Critical
temperature at infinite $L$,  $T_c(L=\infty,N_z)$, versus $N_z$.
MC results are shown by points, continuous line is the prediction of
Capehart and Fisher, Eq. (\ref{CF}). The agreement is excellent. }
\label{TCINF}
\end{figure}

As mentioned above,  $d_{\mbox{eff}}$ is very close to 2. However, $T_c(L=\infty,N_z)$ increases very fast to reach a value close to $T_c$ of the 3D Ising model ($\simeq 4.51$) at $N_z=13$. This is an interesting point.
In  Ref. \cite{Fisher} Capehart and Fisher define the critical-point shift as
\begin{equation}
\varepsilon(N_z)=\left[ T_c(L=\infty,N_z)-T_c(3D)\right]/T_c(3D).
\end{equation}
They showed that

\begin{equation}\label{CF}
\varepsilon(N_z)\approx \frac{b}{N_z^{1/\nu}}[1+a/N_z],
\end{equation}
where $\nu=0.6289$ (3D value). Using  $T_c(3D)=4.51$,  we fit the
above formula with $T_c(L=\infty,N_z)$ taken from Table
\ref{tab:criexp}, we obtain $a=-1.37572$ and $b=-1.92629$.  
Our results and the fitted curve are shown in Fig. \ref{TCINF}. Note
that the correction factor $[1+a/N_z] $ is necessary to obtain a good fit
for small $N_z$.  The prediction of Capehart and Fisher is  verified by our result.

%Fig13

If the cross-over dimension raised by Capehart and Fisher \cite{Fisher} is identified with the effective dimension between 2 and 3, then the hyperscaling relation involving the space dimension $d$ cannot be satisfied in the case of thin films.  Let us take the case  $N_z=13$,   we have $d\nu=2\times 0.9692=1.9384$, while $2-\alpha=2-0.00226=1.99774$ far from the $d\nu$ value.  The hyperscaling relation is thus violated.

\begin{figure}
\centering
\includegraphics[width=4 in]{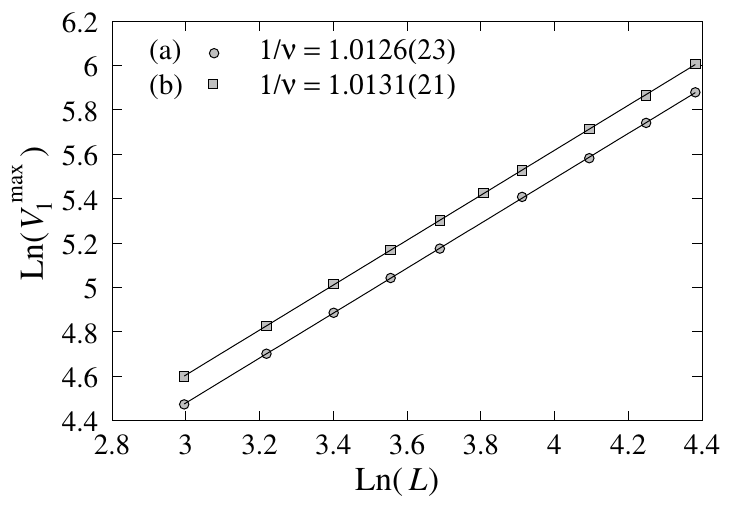}
\caption{Maximum of $V_1$
 versus $ L$ in the $\ln-\ln$ scale for $N_z=5$: (a) without PBC in $z$ direction (b) with PBC in
$z$ direction. The slopes are indicated on  the figure.   } \label{fig:NUZ5}
\end{figure}

\begin{figure}
\centering
\includegraphics[width=4 in]{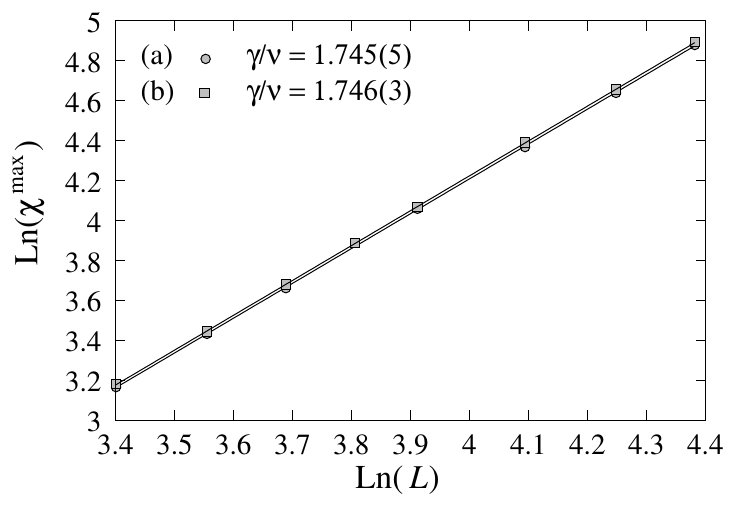}
\caption{$\chi^{\mbox{max}}$ versus $L$ in the $\ln-\ln$
scale for $N_z=5$ (a) without PBC in $z$ direction (b) with PBC in
$z$ direction. The data points of two cases are not distinguishable
in the figure scale.  The slopes are indicated on  the figure.} \label{fig:GAMZ5}
\end{figure}

We show now that the free boundary condition in the $z$ direction gives the same result, within errors,  as the periodic boundary condition (PBC), as far as the critical exponents are concerned. This is	shown in Figs. \ref{fig:NUZ5} and \ref{fig:GAMZ5}.

%Fig14

%Fig15

We  would like to emphasize that  the Rushbrooke inequality $\alpha+2\beta+\gamma \geq  2$ is verified within errors  as an equality for each $N_z$ as seen in Table \ref{tab:criexp}.

\section{Cross-Over from First- to Second-Order Transition with Varying Film Thickness}\label{Crossover}
In this section, we show that the film thickness can alter the nature of the transition. We cite here our work on the cross-over between the first- and second-order transition when the film thickness of a fully frustrated FCC antiferromagnet with Ising spins is decreased to  $\leq 4$ layers (two FCC cells). We used the highly efficient Wang-Landau method \cite{WangLandau,Brown,Schulz,Malakis,NgoSAFTH,NgoSAFTXY,NgoFFSCXY} to detect the thickness where the first-order transition becomes a second-order one.  

\begin{figure}
\centering
\includegraphics[width=3.5 in]{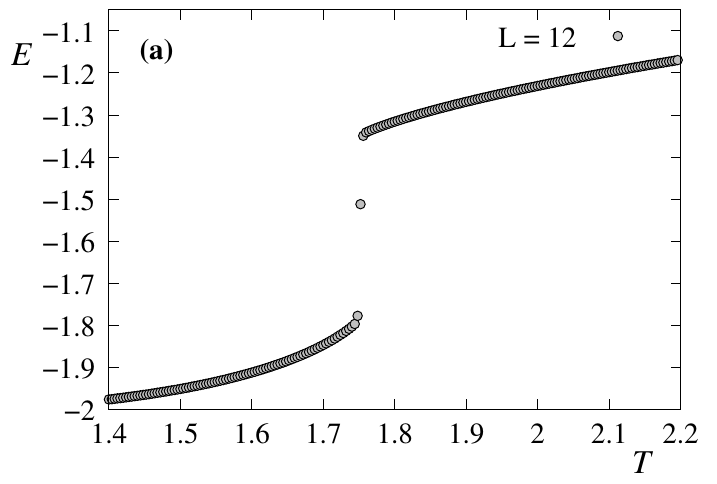}
\includegraphics[width=3.5 in]{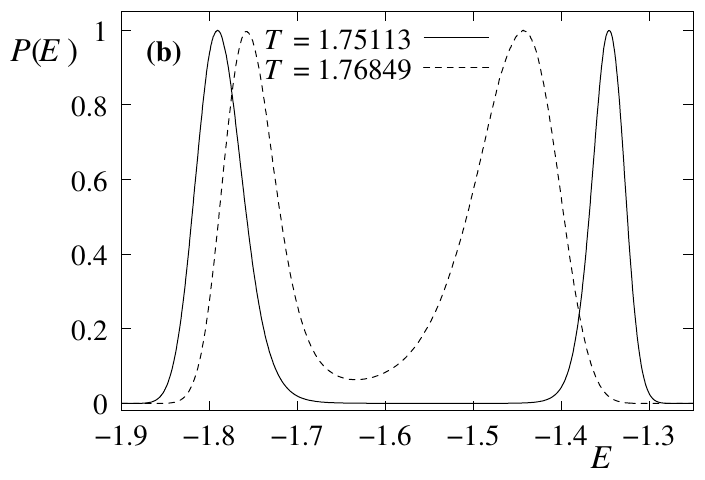} 
\caption{(a) Energy of the bulk case  vs $T$
for $L\times L\times L=12^3$ FCC cells, i. E. the number of spins is $4L^3$; (b) Energy histogram
 with periodic boundary conditions in all three directions (continuous  line) and without PBC (dotted line) in $z$ direction. The histogram was recorded at
the transition temperature $T_c$ for each case (indicated on the figure).} \label{fig:FCC12PE}
\end{figure}

Figure \ref{fig:FCC12PE} shows the strong first-order character of the transition with an energy discontinuity in the bulk case.

%Fig16
\begin{figure}
\centering
\includegraphics[width=4 in]{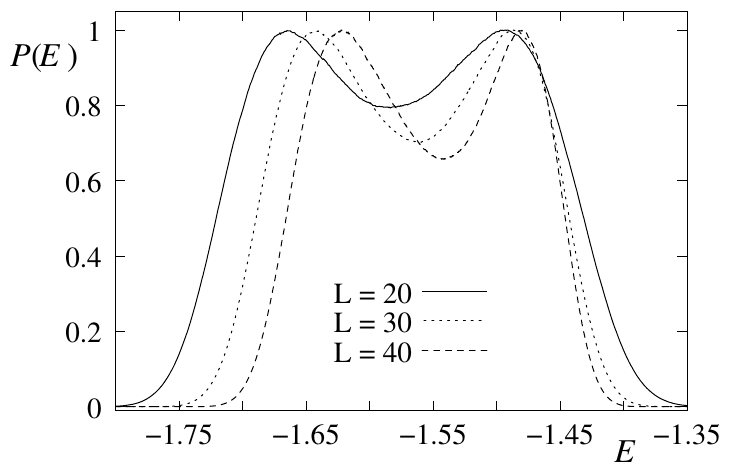}
\caption{Energy histogram
for $L=20,\ 30,\ 40$ with film thickness of 8 atomic layers at $T=1.8218, \ 1.8223, \ 1.8227$, respectively.} \label{fig:Z4PE}
\end{figure}

In the case of a thin film composed of 8 layers ($N_z$=4 FCC cells in the $z$ direction),   the first-order character remains as shown in Fig. \ref{fig:Z4PE} with a double-peak structure.

%Fig17
\begin{figure}
\centering
\includegraphics[width=3.5 in]{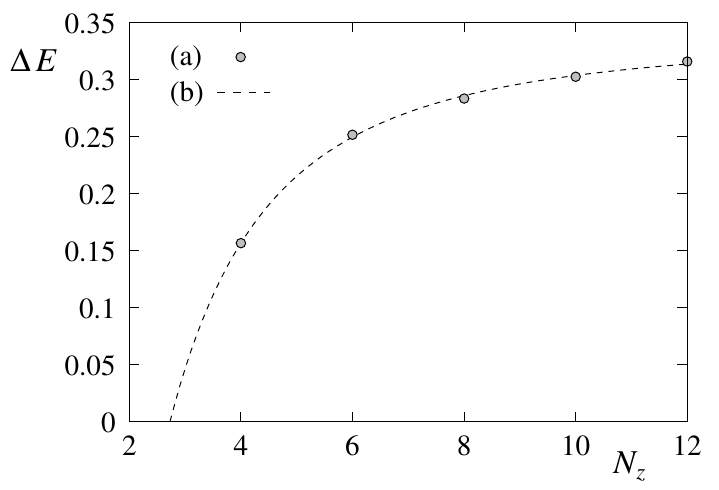}
\caption{The latent heat $\Delta E$ as a function of thickness $L_z=2N_z$ (points are MC results).  The latent heat goes to zero at $N_z=2$, i.e.  at  $L_z=4$ atomic layers. The continuous line  is the fitted function Eq .  (\ref{eq:defit}).}
\label{fig:DENZ}
\end{figure}

When we decrease  the film thickness, the latent heat goes to zero at 4 layers (i. e. $N_z=2$) as shown in Fig. \ref{fig:DENZ}.

%Fig18

We have fitted $\Delta E$  with the following function
\begin{equation}
\label{eq:defit}
\Delta E=A-\frac{B}{N_z^{d-1}}\left[1+\frac{C}{N_z}\right],
\end{equation}
where $d=3$ is the space dimension, $A=0.3370,\ B=3.7068,\ C=-0.8817$.  The second term in the brackets corresponds to a size correction.
As seen in Fig. \ref{fig:DENZ}, the latent heat vanishes at a thickness $L_z=2N_z=4$.  This is verified by our simulations for a 4-layer film: the transition has  a continuous energy across the transition region, even when  $L=150$.

\begin{figure}
\centering
\includegraphics[width=4 in]{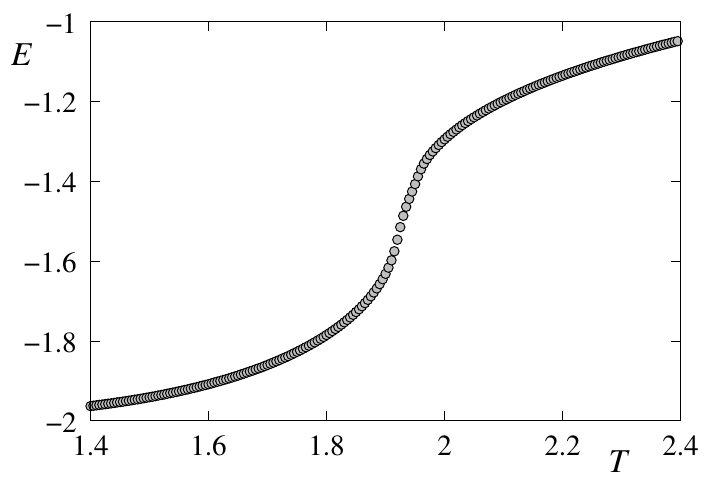}
\caption{Energy  versus temperature $T$
for $L=120$ for a 4-layer film.} \label{fig:L120Z2E}
\end{figure}

The energy versus $T$ for $L_z=4$ is shown in Fig. \ref{fig:L120Z2E}.  

%Fig19
\begin{figure}
\centering
\includegraphics[width=4 in]{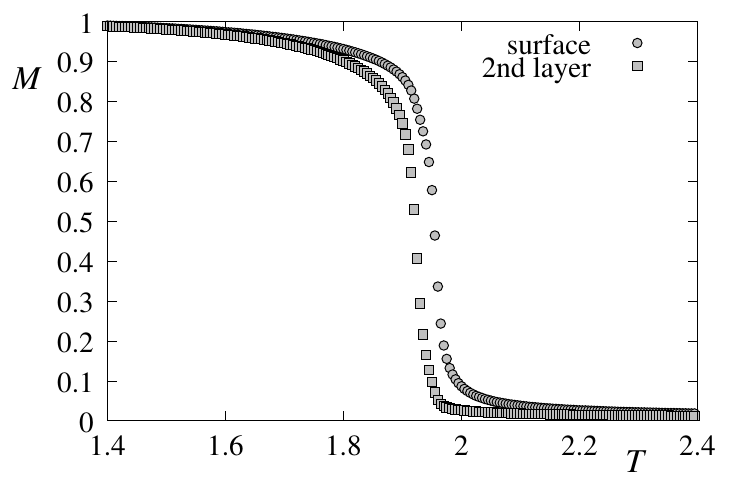}
\caption{Layer magnetizations
for $L=120$ with a 4-layer film: the higher (lower) curve is the surface (beneath) layer magnetization. } \label{fig:L120Z2M}
\end{figure}

As seen in Fig. \ref{fig:L120Z2M},  there are two close transitions: transition of the surface layers at $z=0$ and $z=3/2$,   and that of the beneath  layers at $z=1/2$ and $z=1$ (the lattice constant is taken to be 1).

%Fig20
\begin{figure}
\centering
\includegraphics[width=4 in]{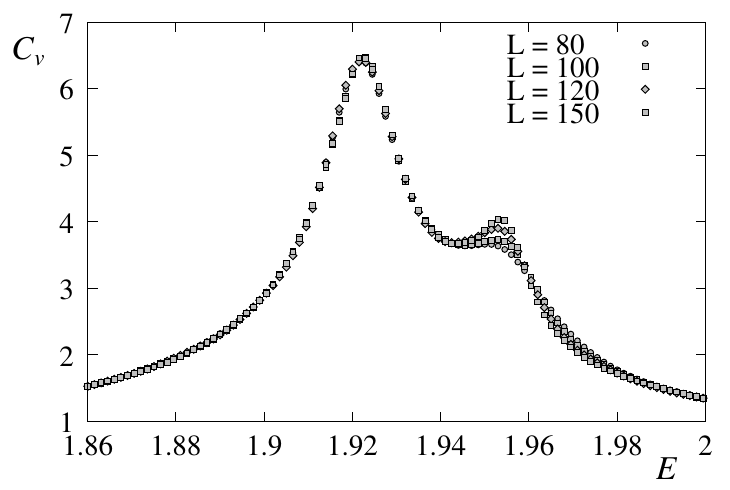}
\caption{Specific heat are
shown for various linear $xy$ plane sizes $L$ versus $T$ for a 4-layer film.} \label{fig:Z2CV}
\end{figure}

\begin{figure}
\centering
\includegraphics[width=4 in]{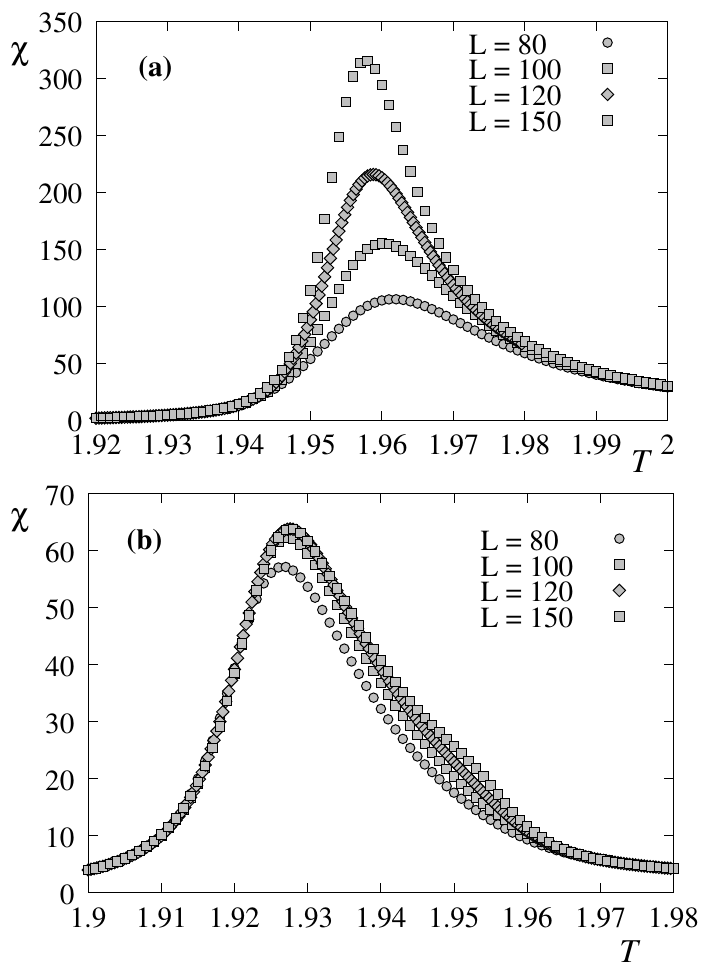}
\caption{Susceptibilities of the first layer (a) and the second layer (b) are
shown for various sizes $L$, versus $T$ in a 4-layer film.} \label{fig:Z2X12}
\end{figure}

\begin{figure}
\centering
\includegraphics[width=4 in]{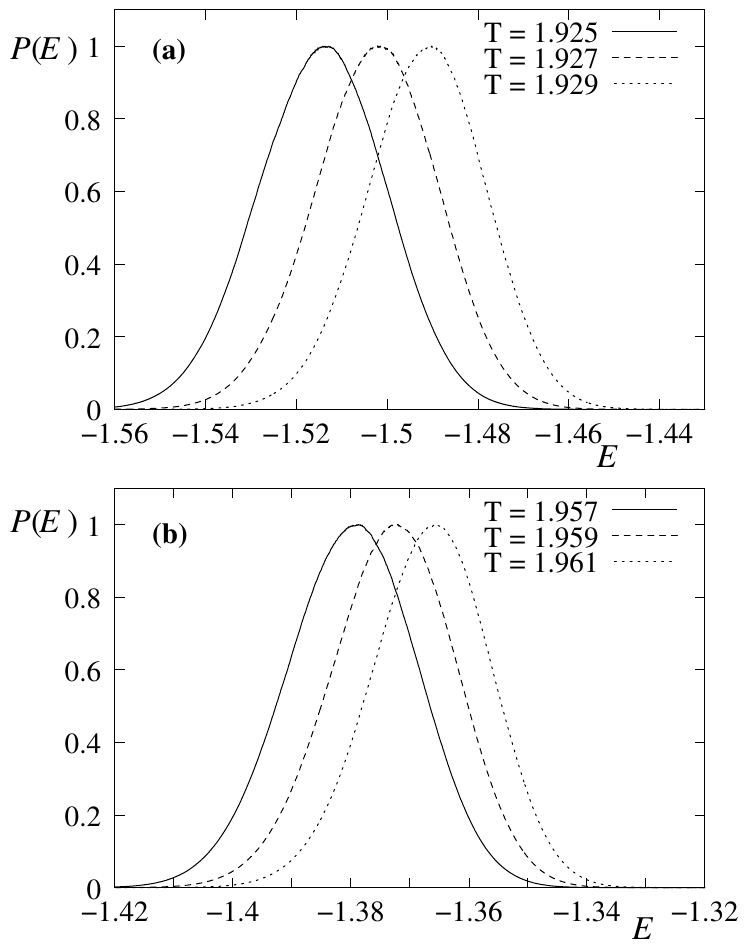}
\caption{Energy histograms recorded 
at temperatures
(indicated on the figure) corresponding to the  the first (a) and second (b) peaks observed in the specific heat, for $L=120$ with 4-layer film thickness $L_z=4$. } \label{fig:L120Z2PE}
\end{figure}

The surface layer has larger magnetization than that of the second layer unlike the non-frustrated case shown in the previous section.  One can explain this by noting that due
to the lack of neighbors, surface spins are less frustrated than the interior
spins, making them more stable than the interior spins.
 This has been found  at the surface of the frustrated helimagnetic film \cite {Diep2015}. 
In order to find the nature of these transitions, using the Wang-Landau technique we study the finite-size effects which are shown in
in Figs. \ref{fig:Z2CV} and \ref{fig:Z2X12}.  The first peak at $T_1\simeq 1.927$  corresponds to the vanishing of the second-layer magnetization, it does not depend on the lattice size,  while the second peak
at $T_2\simeq 1.959$, corresponding to the disordering of the two surface layers, it  depends on $L$.  The histograms   shown in Fig. \ref{fig:L120Z2PE} are taken at and near the transition temperatures show a Gaussian distribution indicating a non first-order transition.

%Fig21
%Fig22

%Fig23
\begin{figure}
\centering
\includegraphics[width=4 in]{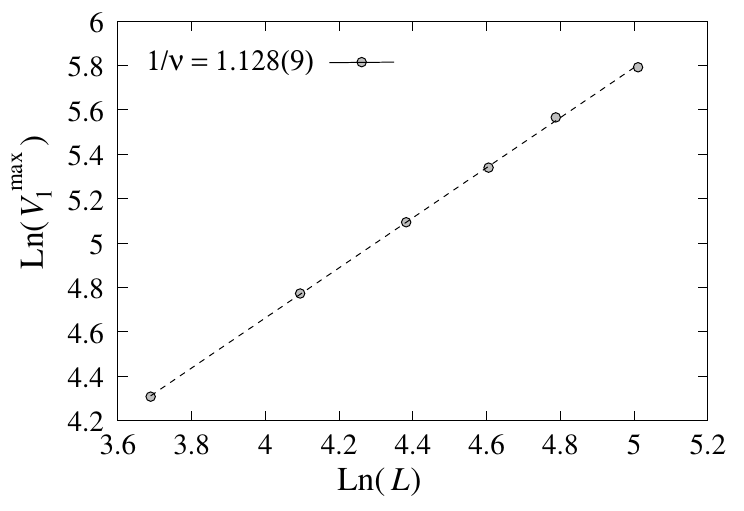}
\caption{The maximum value of $V_1$ as a function of
$L$ in the $\ln-\ln$ scale. The slope of this straight line
gives $1/\nu$. The value of $\nu$ is $\nu= 0.887 \pm 0.009$.}
\label{fig:Z2NU}
\end{figure}

\begin{figure}
\centering
\includegraphics[width=4 in]{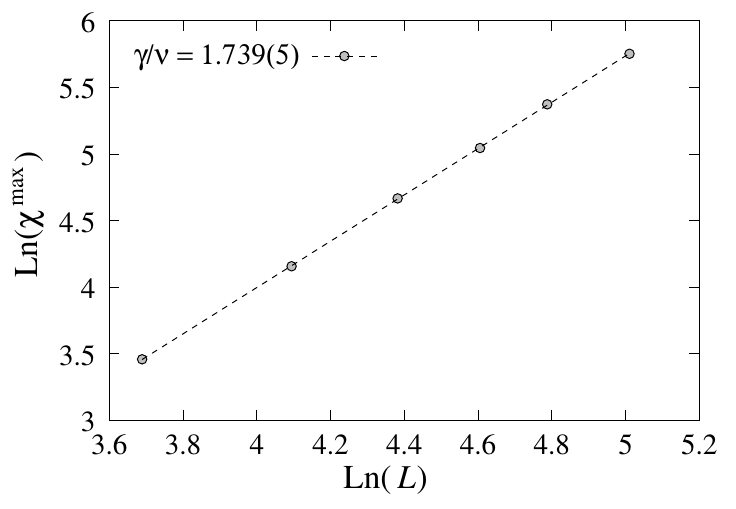}
\caption{The maximum
of the susceptibility  $\chi^{\max}$ as a function of $L$ in the
$\ln-\ln$ scale. The slope of this straight line
gives $\gamma/\nu$.   The value of $\gamma$ is $\gamma =1.542\pm 0.005$ using $\nu= 0.887 \pm 0.009$.} \label{fig:Z2GAM}
\end{figure}

The fact that the peak at $T_1$ of the specific heat does not depend on $L$ suggests two scenarios: i) $T_1$ does not correspond to a transition, ii) $T_1$ is a Kosterlitz-Thouless transition.  We are interested here in the size-dependent transition at $T_2$.
Using the multi-histogram technique, we have obtained $\nu= 0.887 \pm 0.009$ and $\gamma =1.542\pm 0.005$ for the case of a 4-layer film (see Figs. \ref{fig:Z2NU} and \ref{fig:Z2GAM}). These values do not correspond neither to 2D nor 3D Ising models $\nu(2D)=1$,
$\gamma(2D)=1.75$,   $\nu(3D)=0.63$, $\gamma(3D)=1.241$.  We can interpret this as a dimension cross-over between $2D$ and $3D$.   Note that the values we have obtained $\nu= 0.887 \pm 0.009$ and $\gamma =1.542\pm 0.005$ belong to a new, unknown universality class.
At  the time of our work \cite{Ngo2009a}, we relied on the hyperscaling relation with $d=2$ to deduce critical exponent $\alpha$ and using the Rushbrooke equality to calculate $\beta$. However, in view of a possible violation of the hyperscaling when $d$ is not the space dimension, we cannot conclude without a direct calculation of $\alpha$ as we have done in the previous section.
%Fig24

%Fig25

\section{Other cases violating the hyperscaling relation ?}\label{Other}
There are cases where the systems have an additional degree of freedom distinct from the order parameter. This is the case of XY spins with a chirality symmetry: while the chiral symmetry can be mapped onto an Ising-like symmetry, the continuous nature of XY spins affects the criticality of the Ising symmetry breaking. 
Another case is a system of Ising spins in which each spin moves around its lattice site, a kind of magneto-elastic coupling.  These two cases have been previously studied \cite{Boubcheur1998,Boubcheur1999,Boubcheur2001}. We briefly review these results below under the view angle of new universality class and the violation or not of the hyperscaling relation below $d_u=4$. 

\subsection{Fully Frustrated XY Square Lattice}\label{SLFFXY}

\begin{figure}
\centering
\includegraphics[width=3 in]{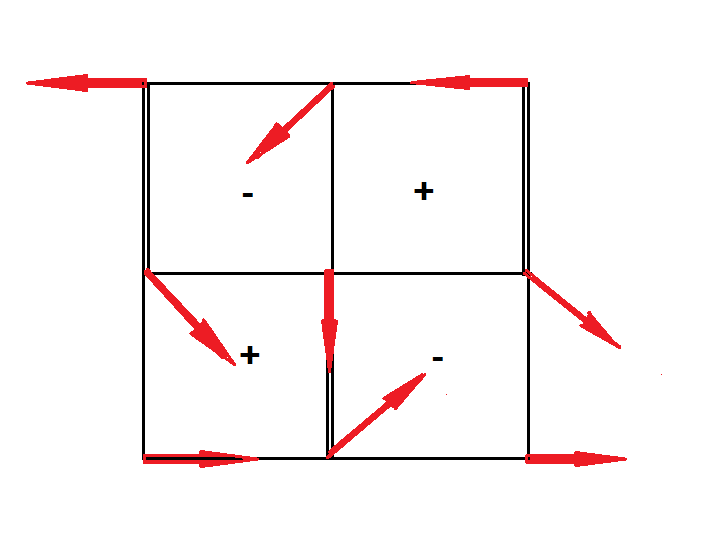}
\caption{The square lattice: single (double) bonds are ferromagnetic (antiferromagnetic) bonds.  The GS spin configuration is shown by red arrows.  The right and left chiralities are denoted by "+" and "-"}.  \label{FFXY}
\end{figure}

The fully frustrated with XY spins on the square lattice  has been intensively studied \cite{Boubcheur1998,Berge,Granato,Lee,Granato2,Olson,Jeon,LeeLee,Ramirez}.  The lattice with the ground-state (GS) spin configuration is shown in Fig. \ref{FFXY} (see \cite{Berge}) where the angles between spins linked by a ferromagnetic (antiferromagnetic) bond is $\pi/4$ ($3\pi/4$) with right and left chiralities.  This GS is equivalent to an Ising model on an antiferromagnetic square lattice. When $T$ increases, one expects a transition of the Ising type. However, at finite $T$, the XY spins fluctuate around their GS orientations shown in Fig. \ref{FFXY}, making the nature of the phase transition more complex as seen below. Note that some authors have claimed that there are two separate transition  of $XY$ and Ising natures \cite{Olson,Jeon,LeeLee,Ramirez}. However, our results \cite{Boubcheur1998} and those of Refs. \cite{Lee,Granato2} show clearly that there is only a single transition of the new criticality called "coupled XY-Ising" universality class.  We show below our results.

%Fig26
\begin{figure}
\centering
\vspace{1cm}
\includegraphics[width=4 in]{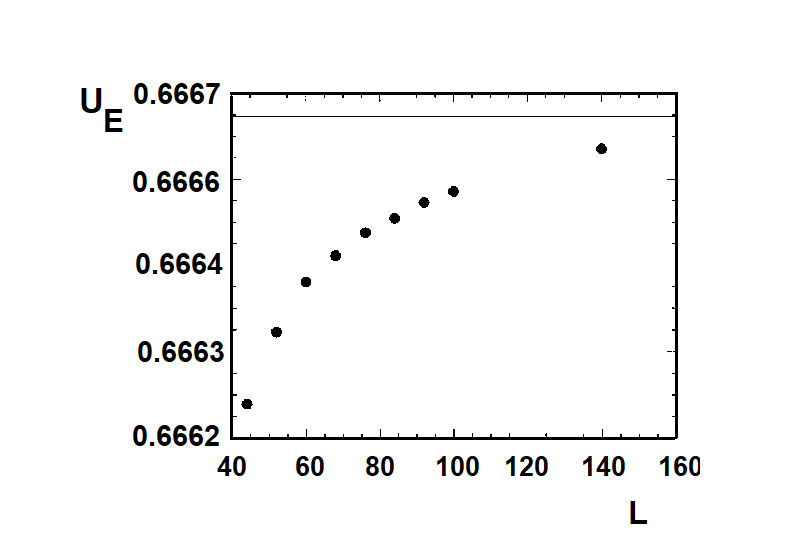}
\caption{The Binder energy cumulant $U_E$ versus $L$ calculated at $T_c(L=\infty)=0.4552(2)$. The horizontal line indicates $U_E(L=\infty)=2/3$. The results show that the transition is of  second order. }  \label{UFFXY}
\end{figure}

In Ref. \cite{Boubcheur1998}, using the highly precise multi-histogram MC simulations described above, we   have obtained the critical exponents $\nu$ and $\gamma$.  We show in Fig. \ref {UFFXY} the Binder energy cumulant $U_E$ \cite{Ferrenberg1,Ferrenberg2}  as a function of $L$: as seen the curve approaches asymtotically $2/3$ from below. This indicates a second-order nature of the transition.
 
%Fig27

\begin{figure}
\centering
\includegraphics[width=4 in]{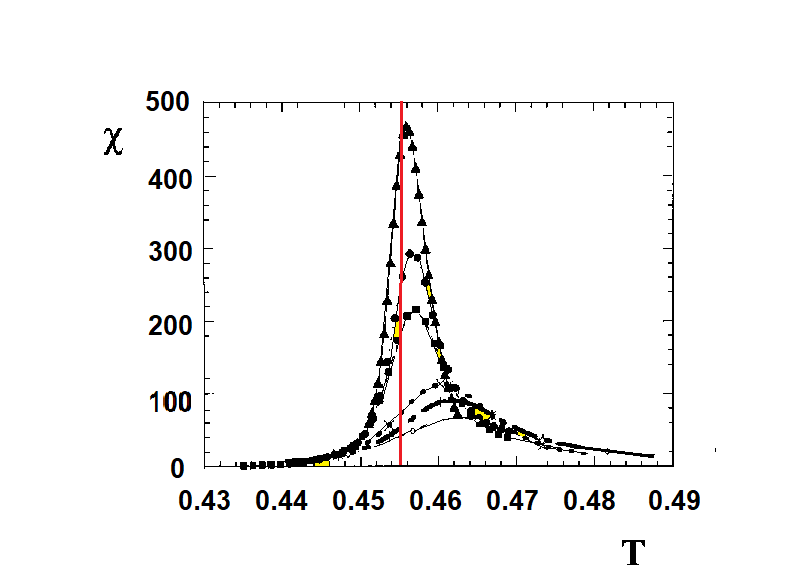}
\caption{Susceptibility $\chi$ versus $T$ for sizes $L=44, 52, 60, 84, 100, 140$ (from lowest to highest curves). The red vertical line is the position of $T_c(L=\infty)$ calculated with Eq. ({TCL}) using $\nu$ obtained  by  Fig. \ref{V1V2FFXY} below. }  \label{ChiFFXY}
\end{figure}

\begin{figure}
\centering
\vspace{0.5cm}
\includegraphics[width=4 in]{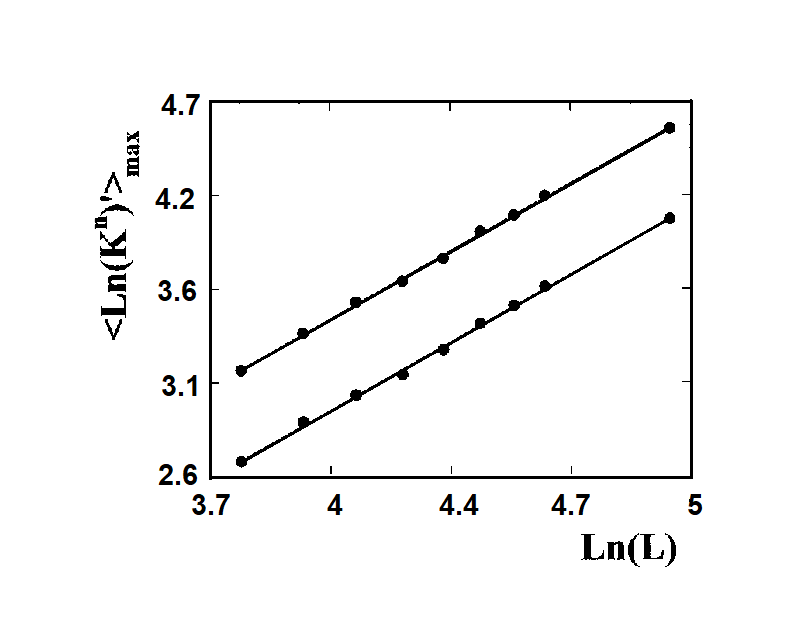}
\caption{The maximum of cumulants $V_1$ and $V_2$ versus size $L=44, 52, 60, 84, 100, 140$ in the $\ln-ln$ scale. The slopes  give the same $1/\nu$. See text for the value of $\nu$.}  \label{V1V2FFXY}
\end{figure}

Let us show the susceptibility versus $T$ for $L=44, 52, 60, 84, 100, 140$ in Fig. \ref{ChiFFXY}. A single peak for each size indicates a single transition as observed above. For $L=140$ the peak is very close to $T_c(L=\infty)=0.45522(2)$ estimated  using Eq. (\ref{TCL}) and $\nu=0.852(2)$ calculated below using the maxima of $V_1$ and $V_2$ shown in Fig. \ref{V1V2FFXY}. The fitting error is less than 0.1\%.  Note that our value is the same as that of Ref. \cite{Granato2} which used  the same multi-histogram MC technique, but differs from those obtained by less efficient methods ($\nu=0.816$ in Ref. \cite{LeeLee} and $\nu=0.889$ in Ref. \cite{Ramirez} for XY transition).

%Fig28

%Fig29

\begin{figure}
\centering
%\vspace{-2cm}
\includegraphics[width=4 in]{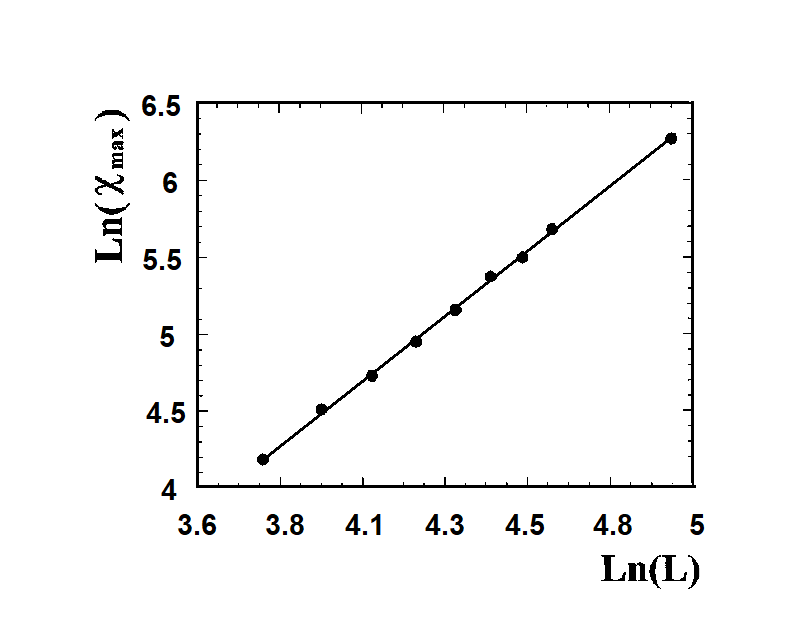}
\caption{The maximum of susceptibility $\chi$ versus size $L=44, 52, 60, 84, 100, 140$ in the $\ln-ln$ scale. The slope indicates the value of $\gamma/\nu$. See text for the value of $\gamma$.}  \label{XmaxFFXY}
\end{figure}

We calculate exponent $\gamma$ using the peak values of $\chi$ for varying $L$. The curve in the $\ln-\ln$ scale  is shown in Fig. \ref{XmaxFFXY}. The slope gives $\gamma/\nu$. Using $\nu=0.852$ we obtain $\gamma=1.531(3)$ which is different from 1.448(24) obtained by Ref. \cite{LeeLee}.

%Fig30

At the time of our work in Ref. \cite{Boubcheur1998}, we did not calculate $\beta$ and $\alpha$ as we have done on thin films presented in the preceding section. We were confident on the hyperscalings $d\nu=\gamma+2\beta$ and $\alpha=2-d\nu$, so we used these relations to calculate $\beta$ and $\alpha$. However, in view of the question whether or not  these relations are valid in particular cases such as the case of thin films shown in section \ref{itf} and the double transition  in the case of fully frustrated square lattice presented here, we would like to check the validity  of the hyperscaling relations. We can use $\nu$ and $\gamma$ obtained here and the value of $2\beta/\nu=0.31(3)$ obtained in Ref. \cite{Granato2}, then we have
$d\nu=2\times 0.852=1.704$ while $\gamma+2\beta=1.531+ 0.31 \times 0.852=1.795$. We see that $d\nu<\gamma+2\beta$ even if we take into account errors of the exponents.  Note that in these estimations we have used the Rushbrooke "equality" which is so far verified within errors in all known cases.  

We note that  it has been found in Ref. \cite{Granato2} by direct FSS that $\alpha/\nu=0.48(7)$. Using $\nu=0.852$ obtained by us and by Ref. \cite{Granato2}, one obtains $\alpha\simeq 0.409$. This yields  $2-\alpha=2-0.409=1.591$ which is not equal to $d\nu=2\times 0.852=1.704$ (here we do not use the Rushbrooke equality).

The  conclusion for this coupled XY-Ising model based on the high-precision multi-histogram technique of our work and of Ref. \cite{Granato2} is that the hyperscaling relation is violated.

\subsection{Effect of magneto-elastic coupling on criticality}

There are certainly other exotic models which may violate the hyperscaling relation for $d<4$. One of these is the magneto-elastic coupling model that we have studied by using the multi-histogram technique \cite{Boubcheur1999,Boubcheur2001}.  The model consists of atoms on a stacked triangular lattice. Each atom carries an Ising spin and moves around its lattice equilibrium position. There are two kinds of interaction which are distance-dependent: the elastic interaction between atoms and the magnetic interaction between Ising spins. We suppose the following Hamiltonian:

\begin{equation}\label{U0Um}
{\cal H}=U_0\sum_{i,j} J(r_{ij}) + U_m\sum_{i,j} J(r_{ij}) \sigma_i\dot \sigma_j,
\end{equation}
where the first sum is the elastic interaction with amplitude $U_0$, and the second sum expresses the interacting spins with amplitude $U_m$.  The distance-dependence  is supposed to be the Lennard-Jones potential 
\begin{equation}\label{LJPot}
J(r_{ij})=(r_0/r_{ij})^{12}-2(r_0/r_{ij})^{6},
\end{equation}
where $r_0=1$ is the distance at equibrium betwenn NN in the triangular planes and also in the stacking direction., $r_{ij}=r_i-r_j$ is the instantaneous distance beween NN. 

In order to separate the spin disordering from  the melting, we take the ratio $Q=U_0/U_m$ large enough so that the magnetic  transition occurs at low $T$.  The cut-off distance  is taken as $r_c=1.366r_0$. 

Simulations using multiplex histogram technique have been performed. The reader is  referred to Ref. \cite{Boubcheur1999,Boubcheur2001} for details. We just summarize the results in Table \ref{tab:criexp2}.  Note that we have calculated at the time of our work (Ref. \cite{Boubcheur1999,Boubcheur2001}) only $\nu$ and $\gamma$, and we have relied on the hyperscaling relations $d\nu=2-\alpha$  and $\gamma/\nu=2-\eta$ to calculate $\alpha$ and $\eta$ listed in Table \ref{tab:criexp2}.
However, we believe that if $\alpha$ is directly calculated as in section \ref{itf}, the result of $\alpha$ may be different, due to the mixing of elastic and magnetic interactions.  The nature of the transition depends on $Q$: it changes from the Ising nature at large $Q$ to close to the XY universality class as seen  in Table \ref{tab:criexp2}.   In such complex situations, we are not sure that the hyperscaling relations are valid. Further direct calculations of $\alpha$ in the way we did in Ref. \cite{Ngo2009} are necessary to conclude on this point.

\begin{table*}
 \centering
   \begin{tabular}{| r | c | c | c | c | c |}
    \hline
    $Q$ & $\alpha$ & $\beta$ & $\nu$ & $\gamma$ & $ \eta$ \\
\hline 8 & $0.140(1)$ & $0.310(1)$ & $0.620(5)$ & $1.245(5)$ & $-0.01(1)$\\
 5 & $0.141(5)$ & $0.301(5)$ & $0.620(5)$ & $1.259(5)$ & $-0.03(1)$\\
 4 & $0.135(6)$ & $0.308(5)$ & $0.622(5)$ & $1.249(5)$ & $-0.01(1)$\\
3D Ising & $0.1070(a)$ & $0.3265(b)$ & $0.6305(b)$ & $1.2390(b)$ & $0.0370(b)$\\
3 & $-0.024(4)$ & $0.353(5)$ & $0.675(5)$ & $1.314(5)$ & $0.053(5)$\\
3D XY & $-0.0100(a)$ & $0.3455(b)$ & $0.671(b)$ & $1.3150(b)$ & $0.040(b)$\\
 \hline
  \end{tabular}\vspace{0.3cm}
 \caption{Critical exponents obtained by multi-histogram technique for different $Q$. $a$: Results from Ref. \cite{Antonenko}, $b$: Results from Ref. \cite{LeGuillou}}\label{tab:criexp2}
\end{table*}

\section{Concluding Remarks}\label{Concl}
We know that the hyperscaling relation is verified in $d=3$ for Ising, XY and Heisenberg spins (results of highly efficient simulations), and in $d=2$ for Ising spins (exact results).  However, as shown in this review, there are particular cases where the hyperscaling relation $d\nu=2-\alpha$ is violated.
One of these situations is  the case of a magnetic thin film with small thickness.  Another case is the fully  frustrated XY square lattice where we show that there is a single phase transition of a new coupled XY-Ising universality: the results  of Ref. \cite{Granato2} and our results using the same method of simulation (multiple histogram technique) are the same for $\nu$ and $\gamma$. We did not calculated $\beta$, but we believe we should obtain the same $\beta$ obtained in Ref. \cite{Granato2}  in view of the same results obtained for $\nu$ and $\gamma$. The hyperscaling relation is then violated in this case. The case of a system with a magneto-elastic interaction shows a new universality class. The violation, or not, of the hyperscaling relation  in this case needs further verifications.

To conclude, let us emphasize that, in view of the precise values we have obtained, at least for thin films, the hyperscaling relation is not verified. We have also presented evidence of the violation of the hyperscaling relation in some other cases.  We believe that more cases should be studied before a general conclusion could be drawn. This explains the question mark in the title of this review.   

Finally, we note that while revising the present paper, we discover that the hyperscaling relation is also violated in the Kuramoto model with the random case (see \cite{Hong}).

\acknowledgments{The authors are grateful to X.-T. Pham-Phu and the late E. H. Boubcheur for participating in some works cited in this paper.}

This research received no external funding.

The authors declare no conflict of interest.

%%%%%%%%%%%%%%%%%%%%%%%%%%%%%%%%%%%%%%%%%%
\end{document}